\documentclass[
    sort&compress,final            
  ]
  {aipproc}

\layoutstyle{6x9}

\def \BEA { \begin{eqnarray}}
\def \EEA {\end{eqnarray}}
\def \BE {\begin{equation}}
\def \EE {\end{equation}}
\def \kS {{\cal H}}
\def \KS {{KS }}

\newtheorem{proposition}{Proposition}

\def \hbm #1 {\mbox{\boldmath{$\hat m^{(#1)}$}}}

\newcommand{\be}{\begin{equation}}
\newcommand{\ee}{\end{equation}}
\newcommand{\beqn}{\begin{eqnarray}}
\newcommand{\eeqn}{\end{eqnarray}}

\newcommand{\ba}{\begin{array}}
\newcommand{\ea}{\end{array}}
\newcommand{\pp}{{\it pp\,}-}

\begin{document}

\title{On Kerr-Schild spacetimes in higher dimensions\footnote{Proceedings of the Spanish Relativity Meeting 2008, Salamanca, September 15-19, 2008 (http://www.usal.es/ere2008/)}}


\author{M. Ortaggio}{
  address={Institute of Mathematics, Academy of Sciences of the Czech Republic \\ \v Zitn\' a 25, 115 67 Prague 1, Czech Republic}
}

\author{V. Pravda}{
  address={Institute of Mathematics, Academy of Sciences of the Czech Republic \\ \v Zitn\' a 25, 115 67 Prague 1, Czech Republic}
}

\author{A. Pravdov\' a}{
  address={Institute of Mathematics, Academy of Sciences of the Czech Republic \\ \v Zitn\' a 25, 115 67 Prague 1, Czech Republic}
  }

\begin{abstract}
 We summarize main properties of vacuum Kerr-Schild spacetimes in higher dimensions. 
\end{abstract}

\maketitle

\section{Introduction}

Kerr-Schild (KS) spacetimes \cite{KerSch652} possess a rare property of being physically important and yet mathematically tractable. In $n=4$ dimensions they contain important exact vacuum solutions such as the Kerr metric and pp-waves. Vacuum \KS spacetimes are algebraically special and thus, because of the Goldberg-Sachs theorem, the \KS null congruence is geodetic and shearfree. Thanks to this and to the Kerr theorem \cite{DebKerSch69,Penrose67,CoxFla76}, the general $n=4$ vacuum KS solution is in fact known \cite{KerSch652,DebKerSch69,Urbantke72,Stephanibook}. 
In arbitrary higher dimensions, the KS ansatz led to the discovery of rotating vacuum black holes \cite{MyePer86}. Here we will study the general class of $n>4$ \KS spacetimes \cite{OrtPraPra08}. 

Note that in $n>4$ gravity there is no unique generalization of the shearfree condition \cite{RobTra83,HugMas88,NurRob02,Trautman02a,Trautman02b,MasTag08} and, {\em a fortiori}, it is not obvious how to extend the Goldberg-Sachs theorem. In fact, it has been pointed out that this can not be done in the most direct way \cite{MyePer86,FroSto03,Pravdaetal04,OrtPraPra07,PraPraOrt07}. Our results below will suggest a possible weak generalization of the shearfree condition, and a partial extension of the Goldberg-Sachs theorem to $n>4$ (limited to \KS solutions). 
It is worth mentioning that an $n>4$ extension of the Robinson theorem has been proven in even dimensions \cite{HugMas88} assuming a generalization of the shearfree condition (see also \cite{NurRob02,Trautman02a,Trautman02b,MasTag08}) different from ours. The relation to our work will be discussed elsewhere. Let us also recall that properties of KS transformations in arbitrary dimensions have been studied in \cite{ColHilSen00}. This does not overlap significantly with our contribution.

\section{Geometric optics and algebraical properties}

By definition, Kerr-Schild spacetimes in $n\ge 4$ dimensions are metrics of the form
\BE
g_{ab} = \eta_{ab} - 2 \kS k_a k_b ,  \label{Ksmetr}
\EE
where $\eta_{ab}$=diag$(-1,1,....,1)$ is the Minkowski metric,  $\kS$ a scalar function
and $k_a$ a 1-form that is assumed to be null with respect to  $\eta_{ab}$, i.e. $\eta^{ab}k_a k_b=0$ ($\eta^{ab}$ is defined as the inverse of $\eta_{ab}$). Hence $k^a \equiv \eta^{ab} k_b= g^{ab} k_b$, so that $k^a$ is  null also with  respect to $g_{ab}$.

It can be shown that optical properties of $k^a$ in the full KS geometry $g_{ab}$ are inherited from the flat background spacetime $\eta^{ab}$. More specifically, the matrix $L_{ij}$ is the same in both spacetimes, i.e
\BE
L_{ij}\equiv k_{a;b}m^{(i)a}m^{(j)b}=k_{a,b}m^{(i)a}m^{(j)b} ,
\EE
and so are the optical scalars expansion $\theta=L_{ii}/(n-2)$, shear $\sigma^2=L_{(ij)}L_{(ij)}-(n-2)\theta^2$ and twist $\omega^2=L_{[ij]}L_{[ij]}$. Furthermore, $k^a$ is geodetic with respect to $g_{ab}$ iff it is geodetic in $\eta^{ab}$. This geometric condition on  $k^a$ turns out to constraint the possible form of the energy-momentum tensor $T_{ab}$ compatible with $g_{ab}$, i.e.
\begin{proposition}
\label{prop_geod}
 The null vector $k^a$ in the \KS metric~(\ref{Ksmetr}) is geodetic iff $\,{T_{ab} k^a k^b=0}$.
\end{proposition}

Note that this condition is satisfied, e.g., in the case of {\em vacuum} spacetimes, also with a possible cosmological constant, or in the presence of matter fields aligned with the \KS vector $k^a$, such as an aligned Maxwell field or aligned pure radiation.

Further computation constrains also the algebraic type of the Weyl tensor
\begin{proposition}
 \label{prop_II}
 If $k^a$ is geodetic, KS spacetimes (\ref{Ksmetr}) are of type II (or more special).
\end{proposition}

From the results of \cite{PraPraOrt07}, it follows that static and (a specific subclass of) stationary spacetimes belonging to the KS class are necessarily of type D, and $k^a$ is a multiple WAND. As a consequence, Myers-Perry black holes must be of type D (cf. also \cite{Hamamotoetal06}). By contrast, black rings do not admit a \KS representation, since they are of type $I_i$ \cite{PraPra05}.   

\section{Vacuum solutions}

In the rest of the paper we will focus on vacuum solutions and, by Proposition~\ref{prop_geod}, $k^a$ will thus be geodetic. Some of the vacuum equations are remarkably simple thanks to the metric ansatz~(\ref{Ksmetr}). In particular, imposing $R_{ij}=0$ we obtain
\BE
 (D\ln\kS)S_{ij}=L_{ik}L_{jk}-(n-2)\theta  S_{ij},  
 \label{vac_ij}
\EE
(where $D \equiv k^a \nabla_a$) and its contraction with $\delta^{ij}$ gives
\BE
 (n-2)\theta (D\ln\kS) 
  =  \sigma^2+\omega^2-(n-2)(n-3)\theta^2.
 \label{Rii=0}
\EE 
The latter involves $\kS$ only when $\theta\neq 0$, so that KS spacetimes naturally split into two families with either $\theta=0$ (non-expanding) and $\theta\not=0$ (expanding).

\subsection{Non-expanding solutions}

It turns out that the vacuum \KS subfamily $\theta=0$ can be integrated in full generality. Our analysis, combined with the results of 
 \cite{Coleyetal06} (where all vacuum Kundt type N solutions have been given) shows that in arbitrary $n\ge 4$ dimensions 
\begin{proposition}
 The subfamily of Kerr-Schild vacuum spacetimes with a non-expanding \KS congruence $k^a$ coincides with the class of vacuum Kundt solutions of type N.
\label{prop_nonexp}
\end{proposition} 

A simple explicit example of non-expanding \KS solutions is given by \pp waves of type N (cf.~\cite{Coleyetal06} and references therein). But note that, as opposed to the case $n=4$, for $n>4$ not all \pp waves fall into the \KS class, and in fact they can also be of Weyl types different from N (see \cite{OrtPraPra08} for details).

\subsection{Expanding solutions}

The subfamily of expanding solutions is more complex and contains, in particular, Myers-Perry black holes \cite{MyePer86}. When $\theta\neq 0$, from eqs. (\ref{vac_ij}) and (\ref{Rii=0}) one gets
\BE
 L_{ik}L_{jk}=\frac{L_{lk}L_{lk}}{(n-2)\theta}S_{ij} .
 \label{optical_const}
\EE
Remarkably, this equation is independent of the function $\kS$. It is thus a purely geometric condition on the \KS null congruence $k^a$ in the Minkowskian ``background'' $\eta_{ab}$ (an {\em optical constraint}). It is important in proving further properties of expanding solutions.

\subsubsection{Optics}

First, the optical constraint implies $L L^T - L^T L = 0$, i.e. $L$ is a {\it normal} matrix. Combining this with the Ricci identities \cite{OrtPraPra07}, one can prove that there exists a ``canonical'' frame in which the matrix $L_{ij}$ takes a specific block-diagonal form,
with a number $p$ of $2\times 2$ blocks ${\cal L}_{(\mu)}$, and a single diagonal block $\tilde{\cal L}$ of dimension $(n-2-2p)\times(n-2-2p)$. They are given by 
\beqn
 & & {\cal L}_{(\mu)}=\left(\begin {array}{cc} s_{(2\mu)} & A_{2\mu,2\mu+1} \label{blockL} \\
 -A_{2\mu,2\mu+1} & s_{(2\mu)} 
\end {array}
 \right) \qquad (\mu=1,\ldots, p) , \\
  & & s_{(2\mu)}=\frac{r}{r^2+(a^0_{(2\mu)})^2} , \quad A_{2\mu,2\mu+1}=\frac{a^0_{(2\mu)}}{r^2+(a^0_{(2\mu)})^2} , \label{s_A} \\
  & &  \tilde{\cal L}=\frac{1}{r}\mbox{diag}(\underbrace{1,\ldots,1}_{(m-2p)},\underbrace{0,\ldots,0}_{(n-2-m)}) , 
 \label{diagonal}
\eeqn
with $0\le 2p\le m\le n-2$. (The integer $m\ge 2$ is the rank of $L_{ij}$. From now on, a superscript (or subscript) index $0$ denotes quantities independent of $r$, which is an affine parameter along $k^a$.) The above special properties of the matrix $L_{ij}$ can be viewed as a ``generalization'' of the shearfree condition and considered in a weak formulation of the Goldberg-Sachs theorem in $n>4$ dimensions, restricted to \KS solutions \cite{OrtPraPra08}.

\subsubsection{Singularities}

Together with the Einstein equation~(\ref{Rii=0}), eqs.~(\ref{blockL})--(\ref{diagonal}) in turn enable one to fix also the $r$-dependence of $\kS$, i.e. 
\BE
 \kS=\frac{\kS_0}{r^{m-2p-1}}\prod_{\mu=1}^p\frac{1}{r^2+(a^0_{(2\mu)})^2} .
 \label{Hgeneral2}
\EE

The above functional dependence suggests there may be singularities at $r=0$, at least for $2p\neq m$ ($m$ even) and $2p\neq m-1$ ($m$ odd). This singular behaviour can indeed be confirmed by examining the Kretschmann scalar. Singularities may also be present in the special cases $2p=m$ and $2p=m-1$ at ``special points'' with $r=0$ and where some of the $a^0_{(2\mu)}$ vanish. See \cite{OrtPraPra08} for more details and \cite{MyePer86} for a thorough discussion of singularities in the special case of rotating black hole spacetimes.

\subsubsection{Weyl type}

Along with the Bianchi identities \cite{Pravdaetal04}, the optical contraint also imply that expanding vacuum \KS solutions can not be of the type III or N, so that in arbitrary dimension $n\ge 4$ 
\begin{proposition}
 Kerr-Schild vacuum spacetimes with an expanding \KS congruence $k^a$ are of algebraic type II or D.
\label{prop_exp}
\end{proposition} 

\begin{theacknowledgments}
  The authors acknowledge support from research plan No AV0Z10190503 and research grant KJB100190702. M.O. also thanks the conference organizers and the European Network of Theoretical Astroparticle Physics ILIAS/N6 under contract number RII3-CT-2004-506222 for financial support to his participation to the Spanish Relativity Meeting 2008.
\end{theacknowledgments}

%
%
%

\end{document}